\documentclass{phbauth}        
\usepackage{graphicx}

\begin{document}

\begin{frontmatter}

\title{Charge and spin inhomogeneity as a key to the physics
of the high $T_c$ cuprates}

\author[address1]{S. Caprara},
\author[address1]{C. Castellani},
\author[address1]{C. Di Castro\thanksref{thank1}},
\author[address1]{M. Grilli},
\author[address1]{A. Perali}

\address[address1]{Istituto Nazionale di Fisica della Materia and
Dipartimento di Fisica - Universit\`a di Roma ``La Sapienza'',\\
P.le A. Moro 2, I-00185 Roma - Italy}

\thanks[thank1]{Corresponding author. E-mail: \\ carlo.dicastro@roma1.infn.it} 

\begin{abstract}
We present a coherent scenario for the physics of cuprate superconductors,
which is based on a chrge-driven inhomogeneity, i.e. the ``stripe phase''.
We show that spin and charge critical fluctuations near the stripe instability
of strongly correlated electron systems provide an effective interaction
between the quasiparticles, which is strongly momentum, frequency, temperature
and doping dependent. This accounts for the various phenomena occurring in the
overdoped, optimally and underdoped regimes both for the normal and the
superconductive phase.
\end{abstract}


\end{frontmatter}

\section{Introduction}
A main issue in the cuprates is to clarify how these systems
evolve from the antiferromagnetic (AF) insulator at very low
doping to the superconductor and to the anomalous normal
metal at higher doping. How does the repulsion characterizing
the AF behavior evolve into an ``effective attraction'' necessary
for superconductivity? Is this evolution connected with the 
anomalous properties of the normal phase? These anomalous properties manifest 
themselves with the linear-in-temperature behavior of the 
resistivity of the CuO$_2$ planes around optimum doping (doping at which
for each cuprate family there is the maximum superconducting
critical temperature $T_c$) and with the opening of pseudogaps
both in the spin and charge channel in the underdoped materials.
Is the strong pairing mechanism
required to obtain the observed high superconducting
temperatures related to these anomalous behaviors of the
normal phase? 

We will show that one way to accomodate the above puzzle into
a coherent scenario is through charge segregation, which may
manifest itself via local phase separation or incommensurate
charge-density waves (ICDW) and stripe formation. In particular
the formation of hole-poor regions and hole-rich regions
enslaves and extends the AF correlations to doping values much 
larger than the doping at which the N\`eel temperature is zero,
since AF correlations 
survive in the hole-poor regions of the segregated domains.

\section{The charge-segregation scenario for the cuprates}
Let us start with the understanding of the anomalous behavior
of the normal phase at optimal doping. One possible explanation
is that the low dimensionality of these highly anisotropic
systems and their correlated nature are at the origin of the
breakdown of the Fermi liquid. However, the proposal that
the Luttinger liquid, the metallic state which is formed in one dimension,
is also  formed in two dimensions \cite{anderson},
has been strongly questioned. 

The alternative attitude has been to 
accept the Landau theory of normal Fermi liquid as a starting
point. The anomalous properties would then arise as a
consequence of strong scattering processes at low energy between the
quasiparticles. 

One possibility,
which we have been discussing along the years, is that 
a singular effective interaction between the quasiparticles 
is mediated by the critical fluctuations occurring
near a charge instability. In particular, within
a perturbative approach, it can be shown that near 
a gaussian critical point with dynamical critical index $z=2$,
the interaction has the form
\begin{equation}
\Gamma_{eff}(q,\omega)\simeq \tilde{U}-\frac{V_c}{|q-
Q_c|^2+\kappa^2-{\rm i}\gamma
\omega}
\end{equation}
 where $\tilde{U}$ is the vestige of the 
strong bare local hole-hole repulsion characterizing the cuprates,
$V_c$ is the strength of the static effective potential, 
$Q_c$ is the critical wavevector related to the 
ordering periodicity, $\kappa^2\sim \xi^{-2}$ is a mass
term which is related to the inverse of the correlation length 
and provides a measure of the distance
from criticality. The characteristic time scale of the
critical fluctuations is $\gamma$. 
We elaborate here on the proposal
that the relevant instability is controlled by
 an incommensurate charge-density-wave (ICDW) 
quantum critical point (QCP) \cite{CDG,ZCDG}.
If the onset of this instability, i.e. the QCP where $\kappa^2=0$,
is located around optimal doping, then in this region no other
energy scale would be present besides the temperature $T$,
as suggested by the in-plane resisitivity experiments. 
The presence of the QCP near optimal doping naturally divides
the phase diagram according to the general scheme of 
criticality into a nearly ordered, a quantum critical, and 
a quantum disordered region, which correspond to the
underdoped, optimally doped, and overdoped regions
respectively. The nearly ordered region is related to the occurrence of
a phase with spatially modulated
charge distribution
(stripe phase) in the underdoped region of the
cuprates, where the charge ordering becomes more pronounced and mixes
with spin degrees of freedom, which are AF correlated in the hole-poor
regions. Owing to this connection, we shall
indifferently use the ICDW- or Stripe-QCP terminology.

>From the theoretical point of view this charge instability is the
natural outcome of the generic tendency 
of models of strongly correlated electrons with
short-range interactions to phase separate (PS)
into hole-rich and hole-poor regions.
This instability is turned into a frustrated
PS or in an ICDW instability
when long-range Coulomb forces are taken into account to guarantee large-scale
neutrality \cite{CDG,noi,emerykivelson,low}. 
The physics of the ICDW-QCP was first derived within
an infinite-$U$ Hubbard model extended by a Holstein electron-phonon
interaction and a long-range Coulomb potential
\cite{CDG,notaraimondi}. With realistic values of the parameters,
the instability was indeed located around optimal doping
and the form (1) for the effective interaction both in
the particle-hole and in the particle-particle channels was derived. 
Another proposal \cite{emerykivelson} for the mechanism of phase
separation is the magnetic interaction. We believe that
the specific mechanism producing phase separation is rather
immaterial, since the strong correlations are the basic ingredient
leading to a charge segregation, whatever residual interaction (magnetic, 
phononic, repulsion between holes on neighbor Cu and O ions,...) 
is considered.  The relative
role of these additional interactions 
might quite naturally depend on doping.  The stripe phase 
continuously connects the low
and intermediate doping regimes were the expulsion of holes from 
the AF background and the  ICDW instability (where 
non-magnetic effects may cooperate)  are respectively dominant.

>From the experimental point of view, 
the existence of charge-controlled inhomogeneities
in some underdoped or optimally doped cuprates is now established.
There is also compelling evidence for a QCP at sizable doping
as provided by magnetoresistivity measurements in
${\rm La_{2-x}Sr_xCuO_4}$:
An insulator-to-metal transition
is found when the SC phase is suppressed by means of
a pulsed magnetic field \cite{boebinger}.
A clear indication that this 
insulator-to-metal transition is 
driven by some spatial charge ordering is provided by its occurrence 
at a much higher temperature in samples near the  ``magic'' filling
1/8, where commensurability effects 
have repeatedly been reported in related compounds.
When extrapolated to $T=0$
the transition takes place near optimum doping. 
The QCP itself is directly observable only when SC is
suppressed. Therefore it would be more appropriate to refer
to a ``hidden'' QCP. 

The stripe-QCP scenario is schematically summarized in Fig. 1,
where the underdoped, optimally and overdoped regions are
apparent.

At large doping the crossover between the quantum critical and quantum 
disordered regime takes place, marking a different temperature and
doping dependence of the CDW correlation length:
$\xi^{-2} \sim T$ in the QC region and 
$\xi^{-2} \sim (x-x_c)$ in the QD one. The effective interaction
(1) due to the fluctuations mediate strong pairing giving rise to 
(d-wave) gap formation in the Cooper channel. The doping and
temperature dependence of the mass term in the pairing potential
gives rise to a critical temperature, which strongly depends on
doping in the overdoped regime and saturates around optimal doping
\cite{PCDG}. Around this latter point the same effective 
interaction in the particle-hole
channel disrupts the Fermi-liquid behavior accounting for the absence
of any energy scale besides $T$. 
In the underdoped region, the 
$T^*$ crossover temperature marks the proximity to the 
nearly charge-ordered phase. Near and below this
temperature the strong critical CDW fluctuations
tend to open pseudogaps in the quasiparticle spectra 
due to the strong scattering in the particle-hole channel \cite{SBBCDG}
taking place around the $(\pm \pi, 0)$ and $(0,\pm \pi)$ points of the
Brillouin zone. 
While in the optimally doped region the temperature itself
provides a mass term to the effective interaction , here the
singular behavior would occur at the onset temperature for the 
stripe phase $T_{CDW}(x)$. $T_{CDW}(x)$ of the order of $T^*$ introduces
a new scale of energy and 
$\xi^{-2}\sim (T-T_{CDW})$.
Approaching this critical line (with  $T_{CDW}(x)$ 
increasing by lowering the doping $x$)
a much stronger fluctuation potential
arises at finite temperature in the  particle-particle
channel. This  results in strongly coupled pairs even though
superconductivity only sets in at lower  $T_c (x)$ values, when
phase coherence between the pairs is established.

\begin{figure}[btp]
\begin{center}
\leavevmode
\resizebox{0.8\linewidth}{!}
{\includegraphics[bb=0 0 610 636]{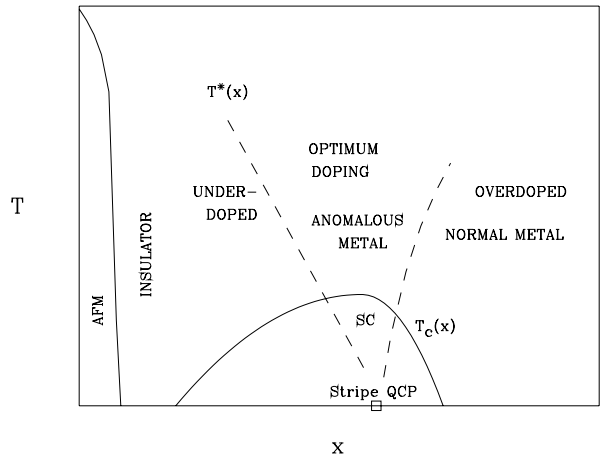}}
\caption{ Schematic phase diagram of the cuprates including the stripe-QCP.
}
\label{figurename}
\end{center}
\end{figure}


\vspace {0.5truecm}

We emphasize here that the most prominent consequence of the presence of
a critical line ending in a
QCP near optimal doping is this strongly singular
effective interaction mediated by the critical charge fluctuations.
This is a distinguished feature of the present scenario,
which provides a more two-dimensional mechanism both for
the pair formation and for the disruption of the Fermi liquid
with respect to other
scenarios, also based on stripe formation, which rely
on the strongly one-dimensional character of the stripes.
 The singlular interaction here described 
has remarkable features: i) It is strongly 
momentum dependent being peaked around the ordering wavevector
related to the specific ordering taking place in the system
(which may vary from system to system and is not
related to the nesting of the Fermi surface); ii) 
it depends strongly on temperature and/or on doping via the
mass term determined by the correlation length. In particular, around 
optimal doping it leaves the temperature as the only relevant energy
scale, whereas in the underdoped regime it introduces the new
scale of energy $T_{CDW}(x)$; iii) it affects
the properties of the system both in the particle-hole and the 
particle-particle channels.

In the following we will present some of the main effects arising
from the singular effective interaction
besides those already generically outlined above. We will first
report the main consequences in the particle-hole channel and 
afterwards in the particle-particle channel.

\section{Stripe fluctuations in the metallic regime:
Effects on the one-particle spectra}
The electronic structure should be directly affected by the
appearence of the anomalous properties described above.
We therefore investigate the effects of stripes 
on the electronic spectra in the metallic phase. Within our approach,
we start from a tight-binding model suitable to describe
the quasiparticles in the cuprates.  The 
quasiparticles interact with the collective (nearly critical)
charge and spin fluctuations. 
In the over and optimally doped regimes, this interaction
can be dealt with perturbatively, whereas
the underdoped phase, where stripes fluctuations are more
pronounced and well formed, requires a different approach,
which at the moment is only partially devised \cite{SBBCDG}.
We consider
an effective interaction between quasiparticles 
of the form (1), which now contains both the charge and the 
(enslaved) spin channels
$\Gamma_{eff}(q,\omega)=-\sum_{i=c,s}
{ V_i/ [\kappa_i^2+\eta_{q-Q_i}
-{\rm i}\gamma_i \omega]
},$
where $\eta_q=2-(\cos q_x+\cos q_y)$, which is
mediated by both
charge ($c$) and spin ($s$) fluctuations. The $q$ dependence
reproduces the $(q-Q_i)^2$ behavior for  $q\approx Q_i$ 
while maintaining the lattice periodicity. The
direction of the critical wavevector $Q_c$ is 
still debated and can be material dependent.

The first-order in perturbation theory  yields an
electron self-energy \cite{CSBDG}, which customarily provides the 
broadening of the quasiparticle peaks, the effective mass
renormalization, and the appearance of incoherent parts in the
single-particle spectra.

The resulting changes of the quasiparticle dispersions, together
with the appearance of (incoherent) dispersing shadow peaks
leads to strong modifications of the spectral densities and of the
Fermi surface , which are in agreement with the observed ARPES 
spectra \cite{saini}.

More recently this analysis was extended \cite{CGCD}
by considering the more realistic
bilayer structure of the Bi2212 compound, where a wealth of ARPES
experiments is available. In particular, the following issue was
addressed:
Although the interplane hopping, as obtained from band calculations,
is sizable ($t_\perp \sim 50 meV$),  the splitting between the two bands
arising from the hybridization between the two planes in the unit cell,
which should be particularly strong near the M
[i.e. $(0, \pm \pi)$ and  $(\pm \pi, 0)$]  points,
is not observed. Even taking into account 
the reduction of $t_\perp$ due to the strong electron-electron
interaction characterizing the cuprates, the bilayer splitting should
still be observable around the M points.
Turning on a moderate coupling between the quasiparticles
and the charge and spin critical fluctuations rapidly dresses and 
drastically reduces the effective interplane hopping, thus washing
out any band-splitting effect and strongly modifying the
Fermi surface shape. This indicates that the puzzling
absence of the band splitting in ARPES spectra of bilayer 
materials can be due to the stripe fluctuations driving
the proliferation of charge and spin collective modes scattering
the quasiparticles. 
This analysis accounts for the good agreement between the
single-layer calculations of the spectra referred above 
with the experimental results on the (two-layer) Bi2212.

\section{Two-gap features}
We want now to explore in more detail how the singular
interactions related to the stripe formation affect
the particle-particle channel. 
The simultaneous presence in Eq. (1)
of a weak momentum independent repulsion together
with a strong attraction at small or intermediate wavevectors
has been shown to favor $d$-wave superconductivity 
\cite{PCDG,GRCDK} within direct calculations in the BCS approximation.
Already within this simplified approach, some non-trivial
features of the superconducting gap were found and
the specific shape of the gap along the fermi surface was not
simply given by the pure $d$-wave form
$\Delta(k)=\Delta_0 [\cos(k_x)-\cos(k_y)]$.
In particular, the gap flattens around the
(1,1) or (1,-1) directions 
when the parameters in Eq. (1) are chosen is such a way that the
effective interaction is strongly peaked around $Q_c$.
Moreover, the gap presents local maxima around the 
k-points on the Fermi surface, which are connected by the
critical wavevectors
(hot spots). Both these features have possibly been identified in
underdoped cuprates. 

Despite these intriguing features of the gap determined in BCS
approximation, some interesting peculiarities of the pairing
effects due to the effective interaction near the stripe singularity line
could have been missed in this approach. In particular, 
 the strong momentum dependence of the effective interaction
gives rise to regions around the hot-spots, where the 
time-reversed ``hot'' states
form nearly local pairs which are tightly bound,
but strongly fluctuating in phase due to the low velocity of the hot points.
These paired states, together with the particle-hole stripe scattering
described in the previous sections can be responsible for the
pseudogaps arising in the underdoped cuprates below $T^*$.
On the other
hand, ``cold'' states far from the hot spots interact more weakly and 
preserve a propagating quasiparticle character. For these states
alone a standard BCS approach would be appropriate and would likely
provide a coherent superconducting state, but at a temperature
much lower than $T^*$. We believe that
the interplay between these states can provide a description of the
pseudogap phenomena in the underdoped regime.
To explore in detail this possibility, a toy model
has been recently investigated \cite{PPVCDG}, where two bands,
say 1 and 2, are present, with a large and small Fermi surface
respectively \cite{nota}. 
While the electrons in the band 1 
represent the cold particles of the cuprates and interact weakly
via a constant attractive interaction $g_{11}$, the hot particles are 
represented by the hot electrons in the band 2, which interact
strongly via $g_{22}$. In this way we schematize electrons in the same band 
interacting with a strongly $q$-dependent potential 
via electrons in two different bands interacting with different
coupling constants. In a BCS mean field approximation
the hot electrons in band 2 would have a high superconducting
critical temperature $T_{c2}^0$, but the smallness of their Fermi surface
leads to a small stiffness $\eta_2$ of the fluctuations, 
which strongly reduce the critical temperature. These fluctuations,
instead do not affect the cold electrons in band 1 having a large Fermi
surface and, therefore, a large stiffness $\eta_1$
(but a low  $T_{c1}^0$). It is 
remarkable that, turning on an interband coupling $g_{12}$,
the stiffness of the hot-pair
fluctuations is increased, but it becomes sizable (of the order of
$\eta_1$) only approaching $T_{c1}^0$.

As a consequence, the coupled 1-2
system acquires a critical temperature value, which is
intermediate between  $T_{c2}^0$ and  $T_{c1}^0$. Thus
the system takes advantage of the strong pairing between the
hot electrons at $T^*\sim T_{c2}^0$ and
reduces the phase fluctuation effects via their coupling
to the colder electrons in band 1 at lower temperature $T_c$.


The above scheme, which we simply described by the introduction of two gaps
finds some experimental support in the different behavior
found for the gap around the M points and along the nodal directions
\cite{panagopoulos,mesot,deutscher,mihailovic}.

In this scenario, the bifurcation of $T^*(x)$ and $T_c(x)$ below a 
doping value of the order of optimum doping and the pseudogap regime
find their natural explanation: i) Around and above optimum doping
the zero-temperature QCP provides the effective interaction which
determines the disruption of Fermi liquid and
the $d$-wave superconductivity; ii) In the underdoped regime, $T_{CDW}(x)$
is increasing by lowering $x$ and shifts to higher temperatures
the effective singular potential, providing strong pairing and
no coherence in the ``hot'' momentum regions and coherence via
the ``colder'' holes.

\begin{ack}
We thank A. A. Varlamov for several helpful discussions.
Part of this work was carried out with the financial support of the
I.N.F.M. - P.R.A. 1996.  
\end{ack}


\end{document}